\documentstyle[12pt]{article}
\global\parskip 6pt

\normalsize

\def\br{\begin{thebibliography}{abc}}
\def\er{\end{thebibliography}}

\def\be{\begin{equation}}
\def\ee{\end{equation}}
\def\bea{\begin{eqnarray}}
\def\eea{\end{eqnarray}}

\begin{document}
\begin{flushright}
\hfill{SINP-TNP/02-06}\\
\end{flushright}
\vspace*{1cm}
\thispagestyle{empty}
\centerline{\large\bf Black Hole Decay as Geodesic Motion }
\bigskip
\begin{center}
Kumar S. Gupta\footnote{Email: gupta@theory.saha.ernet.in}\\
% On leave
%from: School of Mathematics, Trinity College Dublin, Ireland}\\
{\em Theory Division\\
Saha Institute of Nuclear Physics\\
1/AF Bidhannagar\\
Calcutta - 700064, India}\\
\vspace*{.5cm}
Siddhartha Sen\footnote{Email: sen@maths.tcd.ie}\\
{\em School of Mathematics\\
Trinity College\\
Dublin , Ireland}\\
\vspace*{.1cm}
and\\
\vspace*{.1cm}
{\em Department of Theoretical Physics}\\
{\em Indian Association for the Cultivation of Science}\\
{\em Calcutta - 700032, India}\\
\end{center}
\vskip.5cm

\begin{abstract}
We show that a formalism for analyzing the near-horizon conformal symmetry of
Schwarzschild black holes using a scalar field probe is capable of describing
black hole decay. The equation governing black hole decay can be identified
as the geodesic equation in the space of black hole masses. 
This provides a novel geometric interpretation for the decay of black holes. 
Moreover, this approach predicts a precise correction term to the
usual expression for the decay rate of black holes. 
\end{abstract}
\vspace*{.3cm}
\begin{center}
August 2003
\end{center}
\vspace*{1.0cm}
\newpage

The discovery of a conformal group structure in the near-horizon region of a
black hole has led to interesting developments in quantum gravity
\cite{strom,carlip}. The emergence of such a structure is closely related to
the holographic nature of the system \cite{thorn,suss}.
Recently we have shown that by
studying the algebraic structure of an operator that governs the near-horizon 
dynamics of a scalar probe in the background of a massive Schwarzschild black 
hole, the underlying conformal group structure can be explicitly revealed
\cite{ksg1}. Our
analysis was based on the requirement of unitarity of the conformal theory 
and self-adjointness \cite{reed} of the associated near-horizon Hamiltonian
\cite{narn,trg}. 
This led to the introduction of the self-adjoint extension
of the near-horizon Hamiltonian, described by a real parameter
$z$. In this approach, the holographic nature of the black hole is thus
realized by the self-adjoint extension.
It was also shown that for the formalism to be self-consistent, $z$
must be a small positive number. This constraint on $z$ essentially
incorporates the correct boundary conditions required for holography.

In a subsequent paper \cite{ksg2} we showed that further confirmation
of the near-horizon conformal structure emerged when $z$ was
related to the mass $M$ of the Schwarzschild black hole as 
\be
z = \frac{a}{ M^2},
\ee
where $a$ is a nonzero positive constant. For the Schwarzschild background 
it was found that $a = \frac{1}{8}$. For the moment we consider a general
positive value of $a$. It should be stressed that our analysis is valid only
for massive black holes, i.e. for small values of $z$. Once the
identification given by Eqn. (1) is made, it was shown in Ref.
\cite{ksg2} that our approach naturally leads to 
the characteristic logarithmic correction to the Bekenstein-Hawking entropy
\cite{partha,carl1,trg1,danny}.  It is known that such a logarithmic
correction term is universally present in any calculation of black hole
entropy within a conformal field theory framework \cite{danny}.

In this Letter we demonstrate that our formalism \cite{ksg1,ksg2} for
analyzing the near-horizon conformal structure is also
capable of describing black hole decay. For this purpose, we consider the 
space of all self-adjoint extensions which we denote by ${\cal M}$. 
As indicated in Eqn. (1), the self-adjoint extension parameter $z$ is
related to the mass $M$ of the black hole. $M$ can therefore be taken as 
a good coordinate to describe ${\cal M}$.
The central idea here is to study  geodesic motion in the space 
of self-adjoint extensions. We find that within our formalism, there is a 
natural way to give the space ${\cal M}$ a metric.
The geodesic motion in ${\cal M}$ is then calculated using this metric.
We find that the equation for geodesic motion in ${\cal M}$ agrees with the
equation for black hole decay \cite{hawk,wald} for a suitable choice of the 
affine parameter. Once the affine parameter has been so
chosen, our approach predicts a precise correction term in the equation for 
black hole decay arising from the logarithmic correction to the black hole
entropy.

We start by recalling some of the basic results obtained in references 
\cite{ksg1} and \cite{ksg2} which shall be useful for our analysis.
The action ${\cal S}$
for a massless scalar field coupled to a Schwarzschild black hole
of mass $M$
in 3+1 dimensions is given by
\be
{\cal S} = - \frac{1}{2} \int \sqrt{|g|}g^{ij}\partial_i \phi 
\partial_j \phi,
\ee
where the Schwarzschild metric in the spherical polar coordinates has the
form
\be
g_{ij}dx^i dx^j = - (1 - \frac{2M}{r}) dt^2 + (1 - \frac{2M}{r})^{-1}dr^2 + 
r^2 d \Omega^2.
\ee
In the near-horizon limit, 
the corresponding Klein-Gordon operator for the time-independent, zero
frequency and zero 
angular momentum modes of a massless scalar field is given by \cite{trg}
\be
H = - {\frac{d^2}{dx^2}} - {1 \over {4 x^2}},
\ee
where $x = r - 2 M$ is the near-horizon coordinate. In order to study the
quantum properties of $H$, we consider the equation
\be
H \psi = {\cal E} \psi,~~~\psi(0) = 0,
\ee
where ${\cal E}$ is the eigenvalue and $\psi \in L^2[R^+, dr]$ is the 
associated wavefunction. $H$ belongs to the class of unbounded
linear operators on a Hilbert space \cite{reed} and it  
admits a one-parameter family of self-adjoint extensions labelled
by a $U(1)$ parameter $e^{iz}$, where $z$ is real.
The normalized bound state eigenfunction and
eigenvalue of Eqn. (5) are given by \footnote{It may be noted that the
results of our earlier analysis \cite{ksg2} as well as the present one
depends only on the existence of the $n=0$ bound state
of ref. \cite{narn,trg}.} \cite{narn,trg}
\be
\psi (x) =  \sqrt{2 E x} K_0\left( \sqrt{E} x\right)
\ee
and
\be
{\cal E} = -E = - {\exp}\left[\frac{\pi}{2}  {\cot} \frac{z}{2}\right]
\ee
respectively, where $K_{0}$ is the modified Bessel function.
In our formalism, this solution is interpreted as bound
state excitation of the black hole due to the capture of the scalar field.

As mentioned before, the 
requirement of near-horizon conformal symmetry places an 
important constraint on $z$ \cite{ksg1}. 
To see this, consider a band-like region
$\Delta = [x_{0} -\delta/\sqrt{E}, x_{0} + \delta/\sqrt{E}]$,
where $x_0 \sim \frac{1}{\sqrt{E}}$ and
$\delta \sim 0$ is real and positive. When $z > 0$ and satisfies the
condition $z \sim 0$, we see that $x_0 \approx 0$. Under this condition,
$\Delta $ belongs to the near-horizon region of the 
black hole. At any point $x \in \Delta$ the leading behaviour of
the wavefunction is given by \cite{ksg2}
\be
\psi = A \sqrt{E x},
\ee
where $A =\sqrt{2}({\rm ln}2 - \gamma)$ and $\gamma$ is Euler's constant. 
Taking a typical value of $x = x_0$, we can write the wavefunction as 
\be
\psi =  A  E^{\frac{1}{4}} \approx A e^{\frac{\pi}{4 z} }
\ee
for $z \sim 0$. Substituting the value of $z$ from Eqn. (1) in Eqn. (9),
the wavefunction can be written as a function of $M$ as 
\be
\psi (M) =  A e^{c M^2} \equiv A g (M^2).
\ee
The function $g(M^2) = e^{c M^2}$ captures the $M$ dependence of the
wavefunction and $c = \frac{\pi}{4 a}$ is a positive constant. 
For the Schwarzschild background we have $c = 2 \pi $.

The wave function of a system is a natural object to examine 
in order to understand
any symmetry that might be present in the system. To this end, 
consider the set $G \equiv \{g(M^2) | M^2 \in R \}$. The 
elements of $G$ are the functions $g$ defined in Eqn. (10)  
with different values of $M$
corresponding to different elements of $G$. For any two elements 
of $G$ given by
$g(M^2_1) = e^{c M_1^2 }$ and $g(M^2_2) = e^{c M_2^2 }$, 
we can define a composition law as $g(M^2_1) \cdot g(M^2_2) \equiv 
g(M^2_1 + M^2_2) \in G$. Similarly, for any $g(M^2) \in G$, we can define
the inverse element as $g^{-1}(M^2) \equiv g(-M^2) \in G$.
With respect to the composition law defined above, the 
set $G$  has the structure of a continuous abelian group. 

The presence of the continuous abelian group $G$ allows us to describe the 
way the black hole mass changes in a geometric fashion. To do this we 
need to construct a group invariant metric on the space ${\cal M}$.
There is a well known procedure for doing this. We begin
with the observation that the group $G$
has a natural action on the space ${\cal M}$.
Under the action of $G$,
a point $M_0 \in {\cal M}$ transforms as $M_0 \rightarrow e^{c M^2} M_0 \in 
{\cal M}$. $G$ therefore acts as a group of transformations on the space 
${\cal M}$. On a continuous group $G$, the group invariant
metric can be written as \cite{flan}
\be
ds^2 = {\rm Trace}~ (g^{-1} dg )^2.
\ee
In our case, $G$ is abelian and 
using Eqns. (10) and (11), we obtain the expression of the metric on 
${\cal M}$ as 
\be
ds^2 = [d ( {\rm log} g)]^2 = 4 c^2 M^2 (d M)^2 \equiv h_{MM}(dM)^2,
\ee
where $h_{MM}= 4 c^2 M^2$.

We now have all the ingredients to calculate the geodesic equation of motion
in ${\cal M}$. For this purpose, consider a parametrised curve $M(\lambda) \in 
{\cal M}$ where $\lambda \in R$ is taken as the affine parameter.
Using the metric in Eqn. (12), the 
geodesic equation of motion in ${\cal M}$ 
can be written as 
\be
\frac{d^2M}{d \lambda^2} +  \Gamma^M_{MM} 
\left ( \frac{dM}{d \lambda} \right )^2 = 0,
~~~~ {\rm (no~ sum~ over}~ M {\rm )},
\ee
where
\be
\Gamma^M_{MM} = \frac{1}{2}h^{MM} \frac{d h_{MM}}{d M} =
\frac{1}{2} \frac{d}{dM}({\rm log}h_{MM}) = \frac{1}{M}
\ee
and $h^{MM} = h^{-1}_{MM} = (4 c^2)^{-1}M^{-2}$.
In terms of the variable 
$v = \frac{dM}{d \lambda}$, Eqn. (13) can be expressed as
\be
\frac{dv}{dM} + \frac{v}{M}  = 0.
\ee
Integrating Eqn. (15) we get,
\be
v  = \frac{dM}{d \lambda} = \frac{ {\rm constant}}{M}.
\ee
Eqn. (16) describes how the mass of the black hole changes with respect to
the affine parameter $\lambda$ upto an overall undetermined constant. 
It is chosen to be given by $\lambda = M^{-1} t + b$ where $t$ is time
and $b$ is a constant, Eqn. (16) reduces to 
\be
\frac{dM}{dt} = - \frac{k}{M^2},
\ee
where $k >0$ is a constant. The sign of the constant $k$ has been chosen
based on the expectation that the back-reaction
effects will cause the mass of the black hole to decrease in order to
compensate for this energy loss \cite{hawk,wald}. With these
identifications, Eqn. (17) agrees with the standard result of black hole
decay. The assumption of large black hole mass used in
our formalism is a feature present in the usual approach to black hole
decay as well \cite{hawk,wald}. It may be noted that 
the decay rate obtained above is independent of the constant $a$ which
appears in the relation between $z$ and $M$ in Eqn. (1). We thus have a
robust description of the decay of black holes which is independent of the
microscopic details of our formalism. 

The idea outlined above can be applied elsewhere as well.
There is, for instance, a well known example in non-relativistic 
quantum mechanics where
the mass $m$ appears as a parameter in the projective representation of the
Galilean Group \cite{barg}.
Interpreting the set of Galilean transformations as an abelian group 
on the space of masses and following the steps outlined above we get
$\frac{dm}{d \lambda} = \alpha $, where $\alpha $ is a constant. 
Again, with the choice of $\lambda = r t + s$ where $t$ is
time and $r, ~ s$ are constants leads to $\frac{dm}{dt} = \alpha r$.
Choosing $\alpha =0$ we see that this is consistent with conservation of mass.

In ref. \cite{ksg2} we showed 
that the near-horizon conformal structure leads to
a logarithmic correction to the Bekenstein-Hawking entropy. We would now
demonstrate that this logarithmic contribution to the entropy 
generates a corresponding correction term for decay rate of black holes. 
In our formalism, density of states for the black hole 
was related to the modulus square of the 
wavefunction as $ \rho \sim | \psi |^2$ \cite{ksg2}. 
The logarithmic correction to the Bekenstein-Hawking entropy is given
by $- \frac{3}{2} {\rm log} M^2$. A change in the entropy due 
to this term would lead to a corresponding 
change in the density of states. Let $\chi(M)$ denote the 
effective wavefunction associated with this new density of states. In our
formalism, we then have  
\be
| \chi(M)|^2 \sim e^{2cM^2 - \frac{3}{2} {\rm log} M^2},
\ee
which gives 
\be
\chi(M) \sim \frac{1}{M^{\frac{3}{2}}}e^{cM^2}.
\ee
Next we observe that $\chi(M)$ can be written as 
\be
\chi(M) = p(M) g(M^2),
\ee
where $p(M) = \frac{1}{M^{\frac{3}{2}}}$. The set $P = \{p(M) | M \neq 0 \}$
forms an abelian group with respect to the composition law $p(M_1) \cdot
p(M_2) = p (M_1M_2)$ with $p^{-1}(M) = p(M^{-1})$.
The wavefunction $\chi(M)$ thus belongs to the direct
product $P \otimes G$ which is again an abelian group.
Following the analysis presented above, we 
can write the metric on $P \otimes G$ as
\be
ds^2 =  (gp)^{-1} d (gp),
\ee
where $g \in G$ and $p \in P $. In the limit of large black hole mass, 
this metric has the form
\be
ds^2 \approx 4 c^2 M^2 \left [ 1 - \frac{3}{4 c^2  M^2} \right ].
\ee
The corresponding geodesic equation of motion 
gives
\be
\frac{dv}{dM} + v \left [ \frac{1}{M} + \frac{3}{ 4 c M^3} \right ] = 0
\ee
where $v = \frac{dM}{d \lambda}$.
Eqn. (23) can be integrated to give
\be
v M e^{-\frac{3}{8 c  M^2}}  = {\rm constant}.
\ee
Recall that the affine parameter $\lambda$ has already been
identified with $M^{-1} t + b$. With this identification, and for 
large black hole mass, the decay rate of black hole is obtained as 
\be
\frac{dM}{dt} = - k \left [ \frac{1}{M^2}  + \frac{3}{8 c M^4} \right ].
\ee
Note that using the same logic as discussed before, 
the constant in Eqn. (24) has been written as $-k$ with $k > 0$.
Both terms in the r.h.s. of Eqn. (25) contribute to the decay with the same 
sign. The first term on the r.h.s. of Eqn. (25) is identical to what 
was already obtained before through appropriate identification of the affine
parameter $\lambda$. The main result of this Letter is that once $\lambda$
is chosen in the fashion indicated here, the formalism naturally leads
to a precise correction term for the black hole decay rate, determined 
solely by the logarithmic correction to the Bekenstein-Hawking entropy. 
Such a logarithmic term appears universally whenever the Bekenstein-Hawking
entropy is calculated within a conformal field theory framework
\cite{danny}.
It is thus expected that the correction obtained for the black hole decay
should share this same universal property as well.

To summarize we note that the presence of a conformal structure 
in the near-horizon region of a black hole is 
a consequence of the holographic principle. In our formalism, the appropriate 
condition for realizing holography is encoded in the self-adjoint extension 
parameter $z$. The parameter $z$, or equivalently the black hole 
mass $M$, then 
has the natural interpretation as a moduli. It is known that the 
geodesic motion in the moduli space of certain physical system provides an 
appropriate description for the corresponding dynamics \cite{manton}. 
Implementing this idea with an appropriate identification of the
affine parameter is consistent with the standard description of black hole
decay. Moreover, the approach then predicts a correction term to the black
hole decay rate arising from the logarithmic correction to the 
Bekenstein-Hawking entropy. We are thus
led to the surprising conclusion that even the decay of black hole can be
given a novel geometric interpretation within the context of the holographic
principle.

It has been claimed that the near-horizon conformal symmetry is associated
with a large class of black holes in arbitrary dimensions
\cite{carlip,solo}. It seems plausible that probing the near-horizon 
geometry of these black holes would lead to an operator of
the form of $H$, possibly with a different coefficient for the inverse square
term \cite{trg}. It is thus likely that the analysis presented above for the 
massive Schwarzschild black hole could be generalized to include other cases 
as well. 

\newpage

\bibliographystyle{unsrt}

\end{document}